\documentclass[english]{article}
\usepackage[T1]{fontenc}
\usepackage[latin9]{inputenc}
\usepackage{textcomp}
\usepackage{amsmath}
\usepackage{graphicx}
\usepackage{amssymb}

\makeatletter
\newcommand{\lyxaddress}[1]{
\par {\raggedright #1
\vspace{1.4em}
\noindent\par}
}

\makeatother

\makeatother

\usepackage{babel}

\begin{document}

\title{Direct calculation of the attempt frequency of magnetic structures
using the finite element method}

\author{G. Fiedler {[}1{]}, J. Fidler {[}1{]}, J. Lee {[}1{]}, T. Schrefl
{[}2{]}, \\
 R. L. Stamps {[}3{]}, H. B. Braun{[}4{]}, D. Suess {[}1{]}%
\thanks{Corresponding author: Dieter Suess, suess@magnet.atp.tuwien.ac.at%
}}

\maketitle

\lyxaddress{{[}1{]} Institute of Solid State Physics, Vienna University of Technology,
Vienna 1040, Austria}

\lyxaddress{{[}2{]} St. Poelten University of Applied Sciences, St. Poelten 3130,
Austria}

\lyxaddress{{[}3{]} School of Physics, M013, University of Western Australia,
35 Stirling Hwy, Crawley Western Australia 6009, Australia }

\lyxaddress{{[}4{]} UCD School of Physics, University College Dublin, Belfield,
Dublin 4, Ireland}
\begin{abstract}
A numerical implementation of the transition state theory (TST) is
presented which can be used to calculate the attempt frequency $f_{0}$
of arbitrary shaped magnetic nanostructures. The micromagnetic equations
are discretized using the finite element method. The climbing image
nudged elastic band method is used to calculate the saddle point configuration,
which is required for the calculation of $f_{0}$. Excellent agreement
of the implemented numerical model and analytical solutions is obtained
for single domain particles. The developed method is applied to compare
$f_{0}$ for single phase and graded media grains of advanced recording
media. $f_{0}$ is predicted to be comparable if the maximum anisotropy
is the same in these two media types. 
\end{abstract}
\pagebreak{}

\section{Introduction}

A detailed knowledge of the thermal stability is of utmost importance
for magnetic nanostructures for various applications ranging from
hard disc media, MRAM devices to permanent magnets. A well established
tool which originated in chemical rate theory for determining the
thermal stability relies on the transition state theory (TST). In
the TST the thermal stability of a minimum energy state $M_{1}$ is
determined by the application of the Arrhenius-Neel law, $\tau=\frac{1}{f_{0}}e^{\frac{\Delta E}{k_{B}T}}$,
where the energy barrier $\Delta E$ separates the minimum energy
state $M_{1}$ and the saddle point configuration $S_{1}$. The attempt
frequency $f_{0}$ limits the probability for reversal.

The thermal stability, which requires the calculation of the energy
barrier and the attempt frequency was first calculated for single
domain particles in the intermediate to high damping limit (IHD) for
an external field applied exactly parallel to the easy axis\cite{brown_thermal_1963}.
For systems where the symmetry is broken either by an oblique external
field or by an additional anisotropy various works are published giving
analytical formulas for the attempt frequency for different damping
limits. For the IHD limit formulas for the thermal stability can be
found in Ref \cite{smith_classical_1976,brown_jr._thermal_1979,coffey_effect_1998,coffey_thermally_1998}.
For the limit of very low damping a formula for the thermal stability
was derived by Smith et al \cite{smith_classical_1976}. In order
to extend the calculation of the damping limit to all values of the
damping constant Coffey et al. \cite{coffey_crossover_2001}and Déjardin
et al. \cite{dejardin_interpolation_2001}have shown that the Mel'nikov-Meshkov
formalism can be applied to magnetic systems. Analytical formulas
for the intermediate damping regime are given by Garanin et al \cite{garanin_thermally_1999}.

All the previous mentioned work were restricted to reversal via homogeneous
rotation. Inhomogenous states were treated by Braun who calculated
the thermally activated reversal in elongated particles \cite{braun_statistical_1994}.
Besides analytical formulas for the attempt frequency he found that
the thermal stability depends on the domain wall energy in these elongated
particles.

The approach of Braun was extended to soft/hard bilayers, where the
saddle point configuration can be described by a domain wall at the
soft/hard interface by Loxley et al \cite{loxley_theory_2006}.

The paper is structured as follows. In section II details of the implementation
of the transition state theory in the finite element package FEMME
\cite{suess_time_2002}is given. The implementation method is compared
to analytical solutions in section III. An improved method for the
calculation of $f_{0}$ is presented in section IV. Summary and discussion
are given in section V.

\section{Numerical Calculation of the Attempt Frequency}

The numerical implementation is based on Kramers transition state
theory \cite{kramers_brownian_1940}, which Langer transposed and
expanded to multidimensional systems in 1967 \cite{langer_hydrodynamic_1973}.
According to Langers approach the attempt frequency determines the
probability current for the configuration distribution on the energy
surface around the saddle point.

The following steps are required in order to calculate the attempt
frequency by the means of the transition state theory: 
\begin{enumerate}
\item The configuration of the system at the minimum and at the saddle point
given as the values of all degrees of freedom. 
\item The value and curvature of the Gibbs free energy surface at the minimum
and at the saddle point with respect to appropriate canonical variables. 
\item The systems' equation of motion in canonical variables along the energy
surface to describe the dynamics around the saddle point. 
\end{enumerate}
Knowledge of the above properties allows one to calculate the attempt
frequency. This can be written as:

\begin{eqnarray}
f_{0} & = & \frac{\lambda_{+}}{2\pi}\Omega_{0}\label{eqn:f0}\end{eqnarray}

\subsection{Dynamical factor $\lambda_{+}$ from the Landau-Lifshitz-Gilbert
equation}

The dynamical factor $\lambda_{+}$ is obtained by solving the noiseless
linearized equation of motion. In micromagnetics the equation of motion
can be written in the form of the Landau-Lifshitz-Gilbert equation
as,

\begin{eqnarray}
\frac{\partial}{\partial t}\vec{J}=-\frac{\gamma}{1+\alpha^{2}}\vec{J}\times\vec{H_{eff}}-\frac{\alpha}{1+\alpha^{2}}\frac{\gamma}{J_{S}}\vec{J}\times(\vec{J}\times\vec{H_{eff}})\end{eqnarray}

where the magnetic polarization $\vec{J}$ and the effective $\vec{H_{eff}}$
field are continuous functions of space. In the following, we discretize
the continuous equation of motion using the finite element method
using the box scheme \cite{suess_time_2002,gardiner_handbook_1985}.
We can write for the effective field on the node point $i$, \begin{eqnarray}
H_{eff,i}\approx-\frac{1}{V_{i}}\left(\frac{\partial E}{\partial J_{i}}\right)\end{eqnarray}
 with $V_{i}$ being the corresponding volume of spin $i$ and $E$
being the Total Gibbs energy. The structure of the Landau-Lifshitz-Gilbert
equation leads to a constant length of the magnetic polarization $\left|\vec{J}\right|=J_{s}$
. As a consequence the number of independent equation is $2N$, where
$N$ is the number of spins. In order to apply the theory developed
by Langer the system has to be expressed with its canonical variables
\cite{langer_hydrodynamic_1973}. For a system with $N$-spins the
canonical variables on node point $i$ are given by \cite{garcia-palacios_nonlinear_2004,magyari_integrable_1987},

\begin{equation}
\begin{array}{c}
p_{i}=V_{i}J_{s}\cos(\theta_{i})\\
q_{i}=\phi_{i}\end{array}\label{eq:canonical}\end{equation}

where the polar angle angle $\phi_{i}$ and azimuthal angle $\theta_{i}$
are defined as,\begin{eqnarray}
\overrightarrow{J_{i}}=J_{s}\left(\sin\theta_{i}\cos\phi_{i},\sin\theta_{i}\sin\phi_{i},\cos\theta_{i}\right)\end{eqnarray}

The Landau-Lifshitz-Gilbert equation for spin $i$ in canonical variables
reads,

\begin{eqnarray}
\begin{pmatrix}\frac{\partial p_{i}}{\partial t}\\
\frac{\partial q_{i}}{\partial t}\end{pmatrix}=-\frac{\gamma}{J_{s}V_{i}(1+\alpha^{2})}\frac{1}{sin\theta_{i}}\begin{pmatrix}-J_{s}V_{i}\alpha sin^{2}\theta_{i}\frac{\partial E}{\partial\theta_{i}}-J_{s}V_{i}sin\theta_{i}\frac{\partial E}{\partial\phi_{i}}\\
-\frac{\partial E}{\partial\theta_{i}}+\frac{\alpha}{sin\theta_{i}}\frac{\partial E}{\partial\phi_{i}}\end{pmatrix}\label{eqn:llgforlambda}\end{eqnarray}

Let us denote $\eta_{2i-1}=p_{i}$ and $\eta_{2i}=q_{i}$. We can
write the equation of motion in the form, \begin{eqnarray}
\frac{\partial\eta_{2i-1}}{\partial t}=\frac{\partial p_{i}}{\partial t}=:f_{2i-1}(\eta_{1},...,\eta_{2N})\label{eqn:f2iminus1}\end{eqnarray}

\begin{eqnarray}
\frac{\partial\eta_{2i}}{\partial t}=\frac{\partial q_{i}}{\partial t}==:f_{2i}(\eta_{1},...,\eta_{2N})\label{eqn:f2i}\end{eqnarray}

Transition state theory considers the linear dynamics around the saddle
point. Linearizing the right hand side around the saddle point, we
obtain:

\begin{eqnarray}
\begin{pmatrix}\frac{\partial\eta_{1}}{\partial t}\\
\vdots\\
\frac{\partial\eta_{2N}}{\partial t}\end{pmatrix}\approx\left.\begin{pmatrix}f_{1}\\
\vdots\\
f_{2N}\end{pmatrix}\right|_{sp}+\left.\begin{pmatrix}\frac{\partial f_{1}}{\partial\eta_{1}} & \hdots & \frac{\partial f_{1}}{\partial\eta_{2N}}\\
\vdots & \ddots & \vdots\\
\frac{\partial f_{2N}}{\partial\eta_{1}} & \hdots & \frac{\partial f_{2N}}{\partial\eta_{2N}}\end{pmatrix}\right|_{sp}\begin{pmatrix}\eta_{1}-\eta_{sp,1}\\
\vdots\\
\eta_{2N}-\eta_{sp,2N}\end{pmatrix}\label{eqn:lineariseatsaddlepoint}\end{eqnarray}

The first derivatives of the energy with respect to the spin components
at the saddle point is zero, thus the first term on the right hand
side is zero as well. The second term we define as $H_{dyn}$: \begin{eqnarray}
H_{dyn}:=\left.\begin{pmatrix}\frac{\partial f_{1}}{\partial\eta_{1}} & \hdots & \frac{\partial f_{1}}{\partial\eta_{2N}}\\
\vdots & \ddots & \vdots\\
\frac{\partial f_{2N}}{\partial\eta_{1}} & \hdots & \frac{\partial f_{2N}}{\partial\eta_{2N}}\end{pmatrix}\right|_{sp}\label{eqn:hdyn}\end{eqnarray}

The last term we define as vector $\nu_{k}=\eta_{k}-\eta_{sp,k}$.
The time derivative of $\vec{\nu}$ equals the left hand side of Equation
\ref{eqn:lineariseatsaddlepoint}. So Eq. \ref{eqn:lineariseatsaddlepoint}
can be written as:

\begin{eqnarray}
\frac{\partial\vec{\nu}}{\partial t}\approx H_{dyn}\vec{\nu}\end{eqnarray}

This is a linear system of differential equations which has 2N eigenvalues
$\lambda_{k}$ with corresponding eigenvectors $\vec{\nu_{k}}^{0}$.
The solutions are of the form: \begin{eqnarray}
\vec{\nu_{k}}=\vec{\nu_{k}}^{0}e^{\lambda_{k}t}\label{eqn:lambdaplusAnsatz}\end{eqnarray}

With the exception of one eigenvalue, all other eigenvalues of the
matrix $H_{dyn}$ at the saddle point are negative. This single positive
eigenvalue is the $\lambda_{+}$ that we need for the calculation
of the attempt frequency.

For the numerical calculation of $H_{dyn}$, the function $f=f_{i}$
has to be derived with respect to the coordinate $x=\eta_{i}$. This
was calculated numerically using finite differences and a seven point
stencil method.

\begin{eqnarray}
f'(x)\approx\frac{-f(x-3h)+9f(x-2h)-45f(x-h)+45f(x+h)-9f(x+2h)+f(x+3h)}{60h}\label{eqn:7pointstencil}\end{eqnarray}

The step size used in the numerical results was $h=0.005$

\subsection{Statistical factor $\Omega_{0}$ from the Hessian matrix}

The statistical factor $\Omega_{0}$ relates the curvature of the
total Gibbs energy with respect to the canonical variables at the
saddle point and at the minimum. The curvature is obtained by calculating
the second derivative of the energy, \begin{eqnarray}
H:=\left(\begin{array}{ccc}
\frac{\partial h_{1}}{\partial\eta_{1}} & \cdots & \frac{\partial h_{1}}{\partial\eta_{2N-2}}\\
\vdots & \ddots & \vdots\\
\frac{\partial h_{2N-2}}{\partial\eta_{1}} & \cdots & \frac{\partial h_{2N-2}}{\partial\eta_{2N-2}}\end{array}\right)\label{eqn:Hessian_full}\end{eqnarray}

\[
=\left(\begin{array}{ccc}
\frac{\partial^{2}E}{\partial\eta_{1}\partial\eta_{1}} & \cdots & \frac{\partial^{2}E}{\partial\eta_{2N-2}\partial\eta_{1}}\\
\vdots & \ddots & \vdots\\
\frac{\partial^{2}E}{\partial\eta_{1}\partial\eta_{2N-2}} & \cdots & \frac{\partial^{2}E}{\partial\eta_{2N-2}\partial\eta_{2N-2}}\end{array}\right)\]
 $\Omega_{0}$ is the square root of the ratios of the determinants
of the Hessian matrices at the minimum and the saddle point: \begin{eqnarray}
\Omega_{0} & = & \sqrt{\frac{\textup{det}\, H_{min}}{|\textup{det}\, H_{sp}|}}\label{eqn:Omega0}\end{eqnarray}
 It is important to note, that the theory of transitions as developed
by Langer is only valid for canonical variables. Suppose that instead
of the canonical variables$\left(p_{i},q_{i}\right)$, the polar angles
$\left(\theta_{i},\phi_{i}\right)$ are used to describe the system
(see e.g. Ref \cite{schratzberger_validation_2010}) and its derivatives
$\frac{\partial^{2}E}{\partial\theta_{i}\partial\phi_{i}}$. Then
the obtained $f_{0}$ is correct only if the magnetic state at the
minimum and the saddle point fulfills $\theta_{i}=\pi/2$ for all
spins. This prerequisite is, for example, fulfilled for a single domain
particle having the easy axis pointing along the x-axis and an external
field applied along the y-axis as in Ref \cite{schratzberger_validation_2010}.
The same argument applies for the calculation of $\lambda_{+}$.

For the numerical calculation of the derivative the same finite difference
scheme was used as described in the previous section.

\subsection{Climbing Image Nudged Elastic Band Method}

The calculation of the statistical prefactor $\Omega_{0}$ and the
calculation of $\lambda_{+}$ require knowledge of the saddle point
configuration. A very accurate calculation of the saddle point configuration
is essential for the calculation of the attempt frequency. The nudged
elastic band method was used in order to calculate the saddle point
configuration \cite{suess_reliability_2007,dittrich_path_2002}. It
was found that the accuracy of the nudged elastic band method is not
sufficiently high in order to calculate reliably the attempt frequency.
In order to improve the accuracy of the saddle point determination,
a climbing image elastic band method was implemented \cite{henkelman_climbing_2000}.

\section{Validation of the method}

\subsection{Small magnetic cube}

In order to test the numerical calculation of the attempt frequency,
we compared the analytically obtained attempt frequency for a single
domain particle with uniaxial anisotropy to the numerically obtained
one. For the numerical model the following parameters were used: damping
constant $\alpha=1$ , anisotropy constant $K_{1}=1.0MJ/m^{3}$ ,
magnetic saturation polarization $J_{s}=0.5T$ and exchange constant
$A=10pJ/m$. The model size is 0.6 x 0.6 x 0.6 nm$^{3}$. 12 finite
elements were used to discretize the cube. The easy axis of the cube
was assumed to be parallel to the x-axis. The external field was applied
parallel to the y-axis. Results were compared to those obtained using
the analytically derived expression for the attempt frequency given
in Ref. \cite{schratzberger_validation_2010}.

\begin{figure}
\includegraphics[width=10cm]{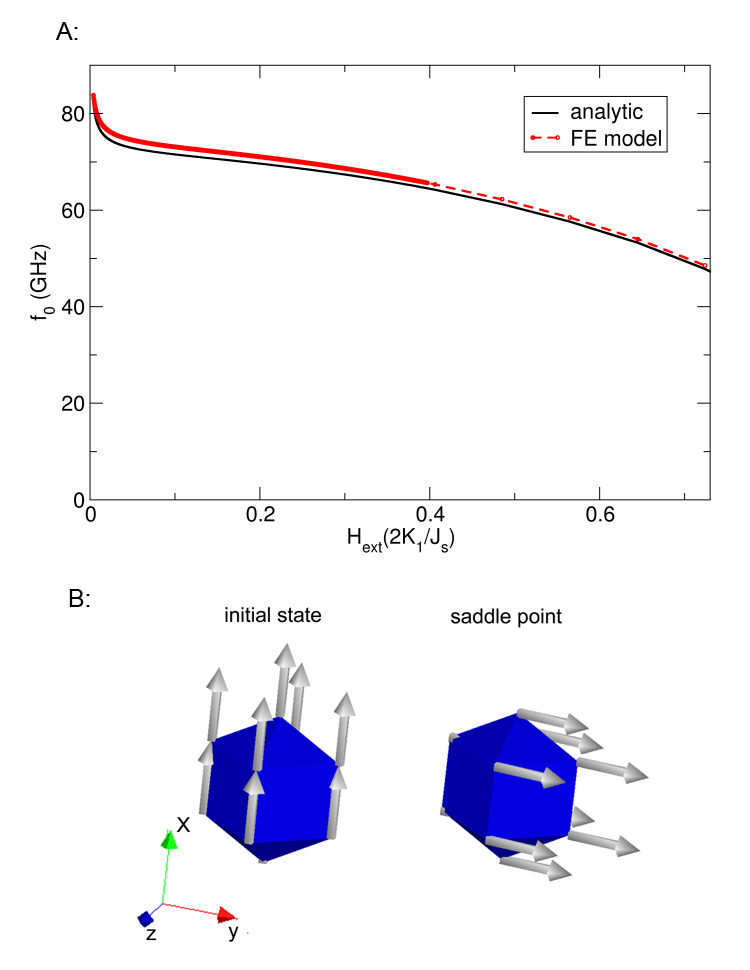}

\caption{\label{fig:analytic}(color online) (A) the numerically calculated
attempt frequency of a small magnetic particle is compared with the
analytical result as a function of the external applied field (H field
applied along y-axis, easy axis parallel to x-axis). (B) The magnetic
state at the minimum (left) and at the saddle point (right) is shown.}

\end{figure}

In Fig. \ref{fig:analytic} (B) two magnetization states of the homogeneous
cube are shown: on the left the initial state is depicted, with magnetization
pointing close to the the easy axis. On the right the magnetization
is shown at the saddle point of the energy, where the magnetization
is perpendicular to the easy axis. The magnetization vectors of all
nodes are shown. Due to the small structure size, all spin vectors
are almost perfectly parallel to each other, indicating that we should
expect a good agreement with the analytic result.

In order to compare the numerical results directly with the analytical
results, the attempt frequency is plotted as a function of the external
field strength as shown in \ref{fig:analytic} (A) . The simulations
agree very well with the analytical solution. It should be noted,
that both the analytical solution and the numerical simulation are
not valid for zero external field, since the expressions for field
lowered symmetry are used. As a consequence the spike-like increase
at zero external field is an artifact and has no physical meaning.

\subsection{Elongated particle}

As a second example we calculate the attempt frequency of an elongated
particle. The simulated geometry of the elongated grain is: length
25\,nm, width 1.6~x~1.6\,nm. This shape is similar to the magnetic
grains used as the recording media in modern hard disks, only slimmer
(die lateral dimension of a grain of a hard disc is about 6~x~6\,nm).
According to the analytical theory of Ref. \cite{loxley_theory_2006}the
cross section area does not influence the attempt frequency. Hence,
in order to save computational time we model the grain of a recording
media with a smaller cross section but we can expect that the results
can be used for realistic grains.

The following parameters were used for the graded media grain example:
damping constant $\alpha=0.2$ , anisotropy constant increasing quadratically
from zero to $K_{1}=3.6MJ/m^{3}$ , magnetic saturation polarization
$J_{s}=0.5T$ and exchange constant $A=10pJ/m$. The material parameters
are the same as in Ref \cite{suess_thermal_2008}.

\begin{figure}
\includegraphics[width=10cm]{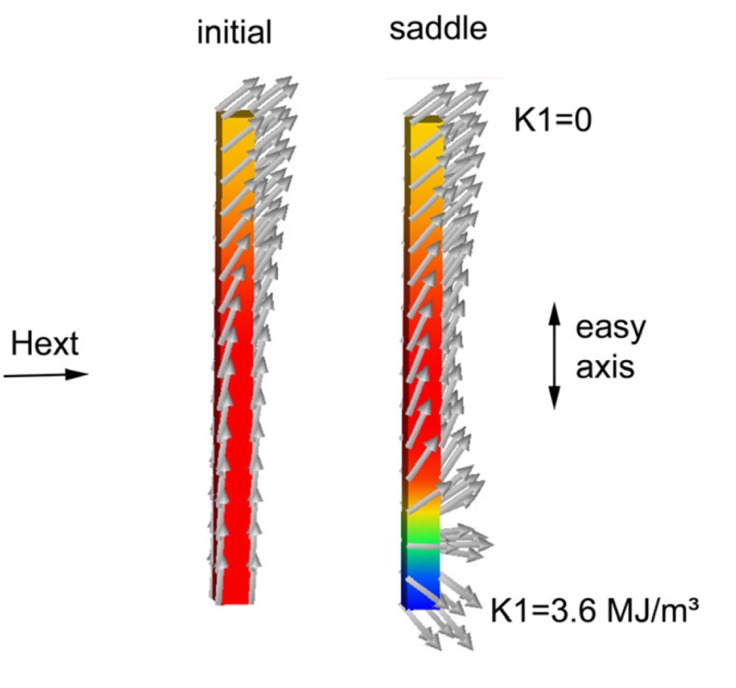}

\caption{\label{fig:graded}(color online) model of the graded media grain
which is soft magnetic at the top and hard magnetic at the bottom.
(left) energy minimum state (right) saddle point configuration.}

\end{figure}

Fig. \ref{fig:graded} shows the simulated grain with quadratically
increasing $K_{1}$ . The directions of the easy-axis and the applied
perpendicular external magnetic field are shown. The magnetic state
at the minimum of the energy is shown on the left side of the figure,
and the saddle point is shown on the right side.

As in the previous section the attempt frequency was calculated as
a function of a perpendicular applied field. In order to verify reliability,
the simulations were repeated with different finite element meshes.

\begin{figure}
\includegraphics[width=10cm]{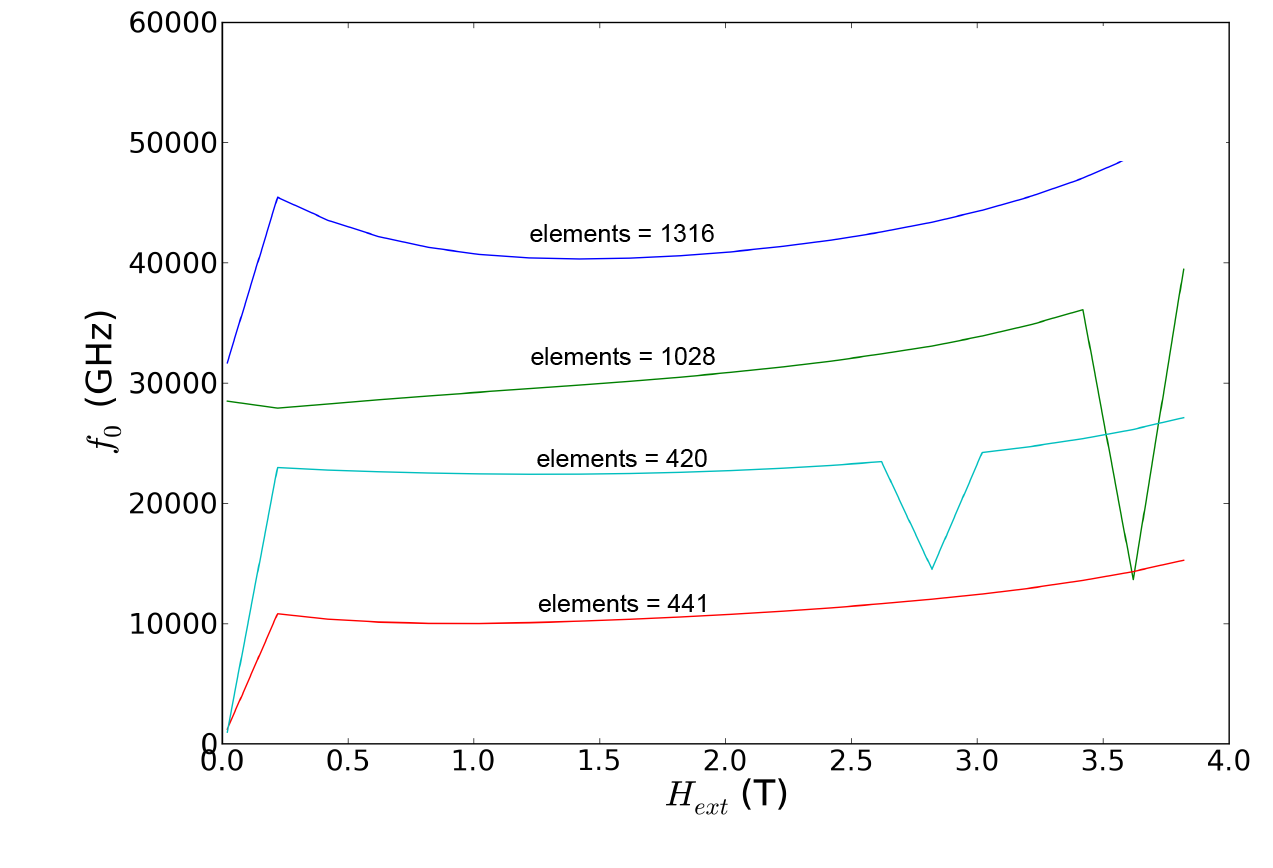}

\caption{\label{fig:mesh_dependance_graded}(color online) The attempt frequency
as a function of the external field strength is shown for a graded
media grain. The external field is applied perpendicular to the easy
axis. The results are shown as a function of the number of finite
elements. }

\end{figure}

Fig. \ref{fig:mesh_dependance_graded} shows the attempt frequency
of four different finite element meshes, with different numbers of
volume elements. As can be seen, the results of the attempt frequency
do not converge to a single value as the finite element mesh size
is decreased. Furthermore the obtained value of the attempt frequency
is one order of magnitude larger than the value reported in Ref \cite{suess_thermal_2008}.
The reason for these discrepancies are discussed in the next section.

\section{Improved method using analytical fits}

The reason for the numerical errors of the attempt frequency can be
found in the calculation of $\Omega_{0}$, which consists of (the
square root of) a ratio of determinants as given by Eq.\ref{eqn:Omega0}.
The calculation of the determinant of the symmetric matrix $H$ equals
the product of all its eigenvalues $\lambda_{1}$, . . . , $\lambda_{2N}$,
where $N$ is the number of finite element node points.

Hence, we can write $\Omega_{0}$ in the form:

\begin{align*}
\Omega_{0} & =\sqrt{\frac{\prod\lambda_{min_{i}}}{|\prod\lambda_{sp_{i}}|}}=\sqrt{\left|\frac{\prod\lambda_{min_{i}}}{\prod\lambda_{sp_{i}}}\right|}\end{align*}
 All eigenvalues of $H_{min}$ at the minimum are positive. At the
saddle point there is only one negative eigenvalue. The number of
eigenvalues increases with increasing number of finite elements. Hence,
small numerical errors in the calculation of the eigenvalues multiply
together to form a considerable total error.

\begin{figure}
\includegraphics[width=10cm]{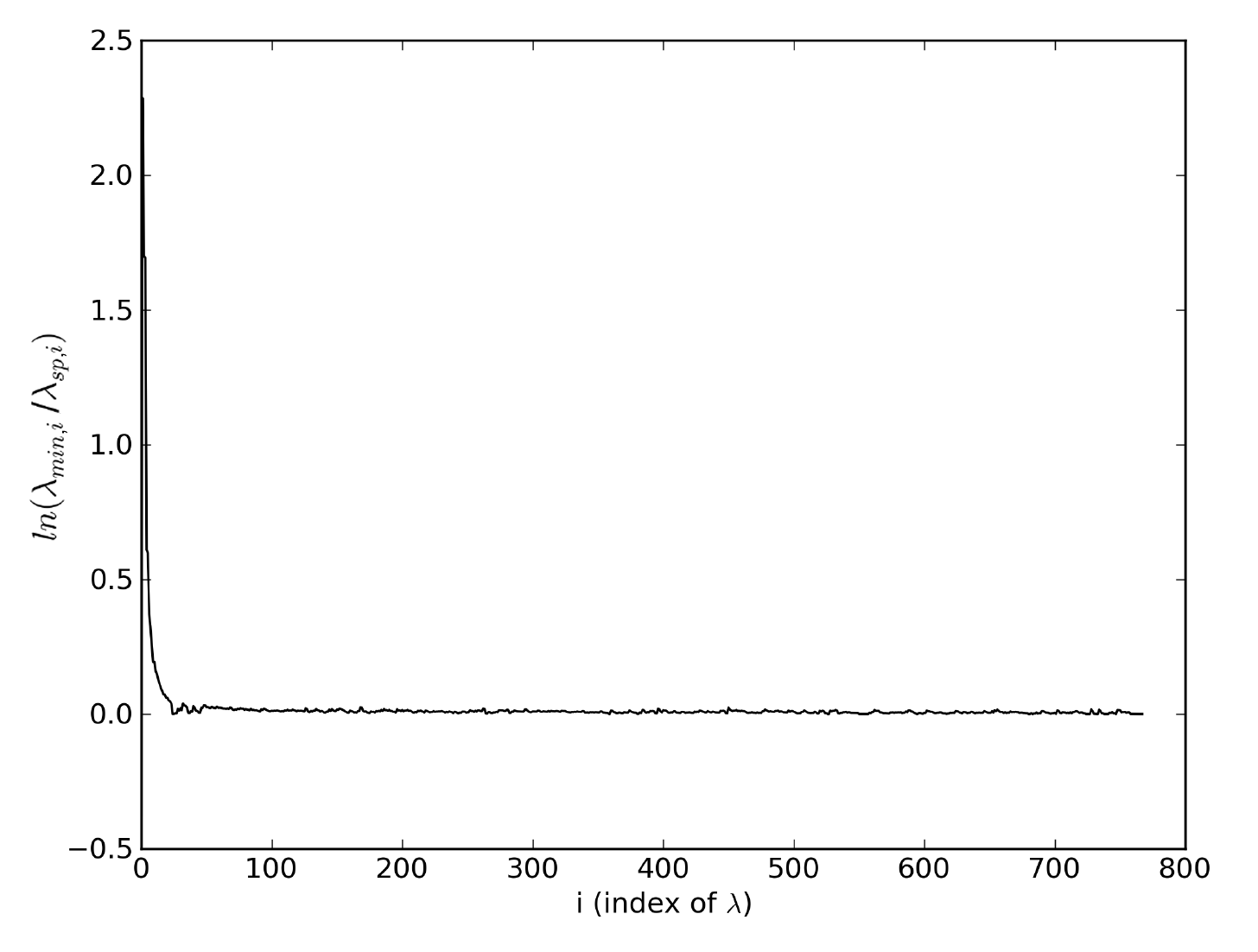}

\caption{\label{fig:Logarithmic-ratios-of}Logarithmic ratios of eigenvalues
as a function of the eigenvalue index i. (number of finite elements=
1316 , H$_{ext}$ is 0.02 T.)}

\end{figure}

Fig. \ref{fig:Logarithmic-ratios-of} shows the ratio of the eigenvalues.
The x-axis represents the index $i$ of the eigenvalues, and the y-axis
represents the natural logarithm of the ratios of eigenvalues, $\lambda_{min,i}/\lambda_{sp,i}$.
In Fig. \ref{fig:Logarithmic-ratios-of} the eigenvalues are sorted
according to modulus, where the eigenvalue with the index $i=1$ has
the smallest modulus and $i=2N$ is the eigenvalue with the largest
modulus. As shown in Fig. \ref{fig:Logarithmic-ratios-of} only ratios
of eigenvalues with small i are not close to or equal to 1. All the
ratios of the eigenvalue with $i>>100$ are very close to one, which
results in $\ln(\lambda_{min,i}/\lambda_{sp,i})\thickapprox0$ for
$i>>100$.

The numerical evidence in Fig. \ref{fig:Logarithmic-ratios-of} illustrate
the insensitivity of the saddle point to the majority of eigenmodes,
and the thermal significance of modes with the smallest eigenvalues.
As noted in Ref \cite{livesey_spin_2006}, a domain wall pinned at
an interface supports a broad range of travelling spin waves that
are essentially unperturbed by the wall structure, and a set of modes
localized to the domain wall. The modes localized to the wall have
the smallest eigenvalues, and represent the most relevant fluctuations
for thermal depinning of the wall. 

Braun has shown that for elongated particles, the logarithm of the
ratio of eigenvalues scales with $\frac{i}{1+i^{2}}$ (see formula
(4.12) of Ref. \cite{braun_statistical_1994}).

\begin{figure}
\includegraphics[width=10cm]{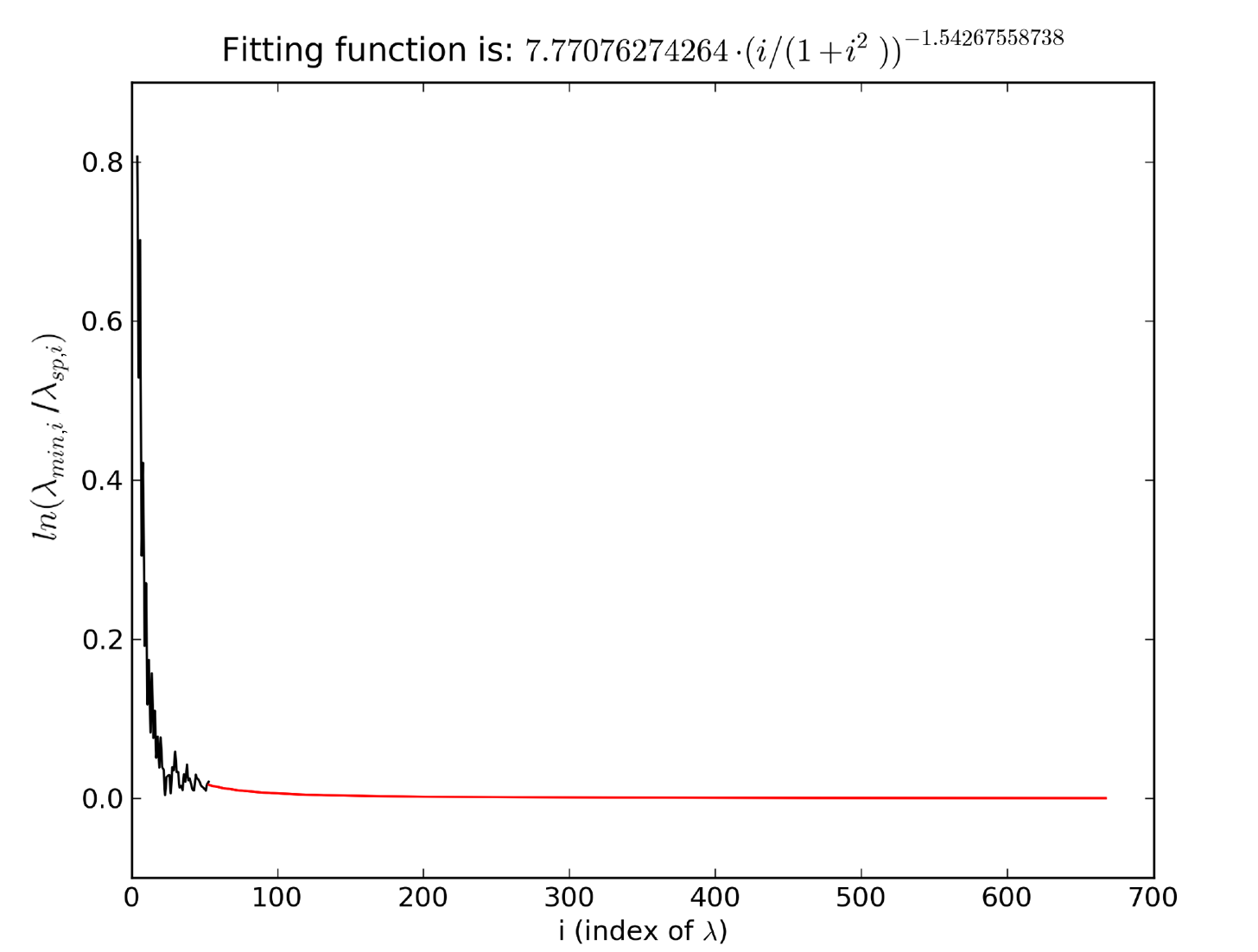}

\caption{\label{fig:fittedfunction}Contributions to calculation of $\Omega_{0}$:
up to index 50 the original data is taken (black line), and from index
50 onwards the values of the fitting function are used (red line).}

\end{figure}

In order to reduce the numerical error which is introduced by the
higher order eigenfrequency, we fit the values of $\ln(\lambda_{min,i}/\lambda_{sp,i})$
to the following function,

\begin{eqnarray}
f(i)=a\cdot\left(\frac{i}{1+i^{2}}\right)^{b}\label{eqn:fitfunction}\end{eqnarray}

, where $a$ and $b$ are parameters determined by a mean square fit
to the numerical data for $5<i<60$. In Fig. \ref{fig:fittedfunction}
the black line represents the ratios of eigenvalues, calculated by
the numerical simulation, and the red line is the fitted function,
which values of parameters $a$ and $b$ are given in the inset. The
calculation of $\Omega_{0}$ is done by calculating $\Omega_{0}=\sqrt{\left|\prod_{i=0}^{c}{\frac{\lambda_{min_{i}}}{\lambda_{sp_{i}}}}\cdot\prod_{i=c+1}^{N}{e^{fit(i)}}\right|}$.
We found for a wide range of c ($10<c<100$) that the value of $\Omega_{0}$
is not significantly influenced by the actual choosen value of the
paramter c.

\begin{figure}
\includegraphics[width=10cm]{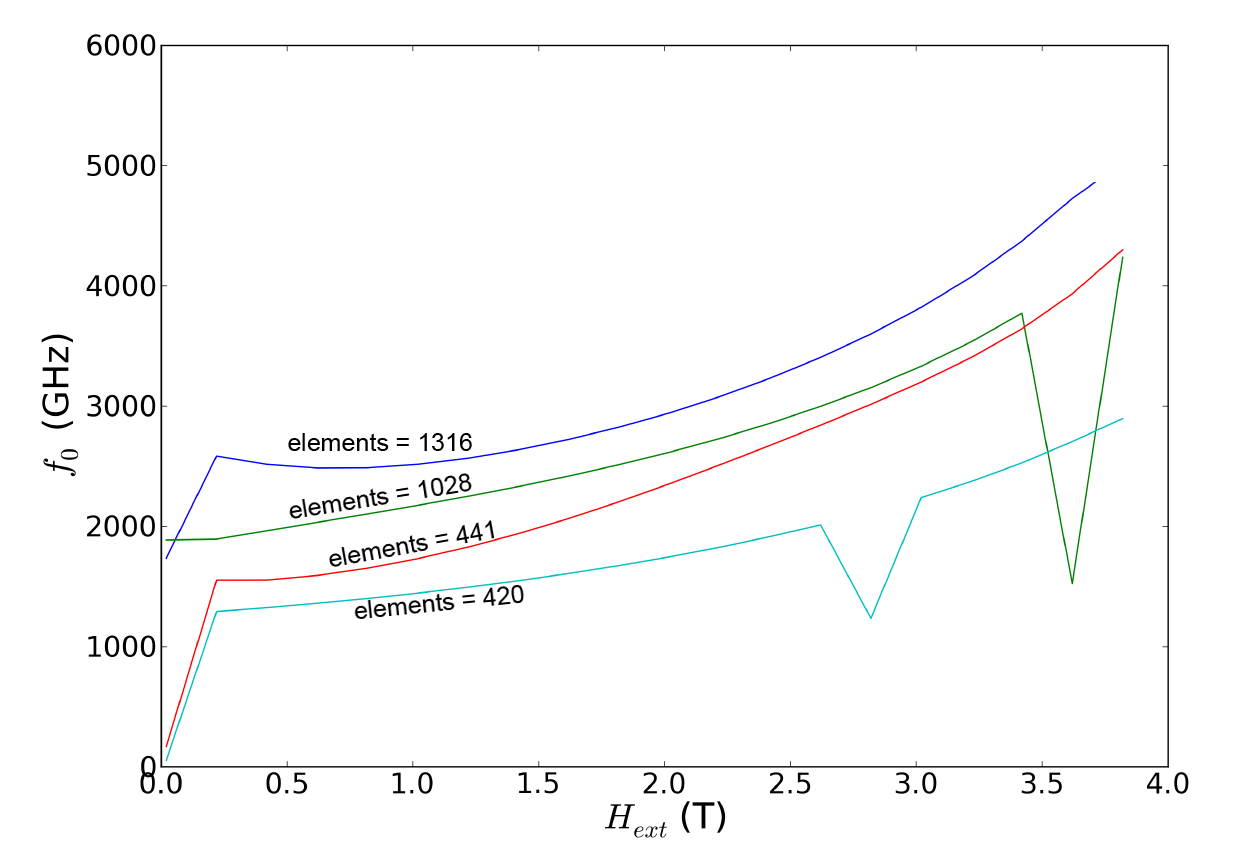}

\caption{\label{fig:graded_fitted}Same as Fig. \ref{fig:mesh_dependance_graded}
except that a fitting function from eigenvalue index 30 onwards is
used for the calculation of $\Omega_{0}$.}

\end{figure}

The calculation of the attempt frequency of the graded media grain
is repeated with the improved method using the fit function. Fig.
\ref{fig:graded_fitted} shows that the mesh dependency of the results
is decreased by this method. If the value of the attempt frequency
is extrapolated for an external field approaching zero we can estimate
the attempt frequency to be in the order of $f_{0}\sim1850GHz\pm650GHz$.
Comparing this estimate of the attempt frequency with the estimate
of the attempt frequency obtained by Langevin dynamic simulations
(In Ref. \cite{suess_thermal_2008} for the same system a value of
$f_{0}\sim1638GHz\pm46GHz$ is estimated), despite the various assumptions
which are used in both methods, a considerable good agreement can
be found.

\begin{figure}
\includegraphics[width=10cm]{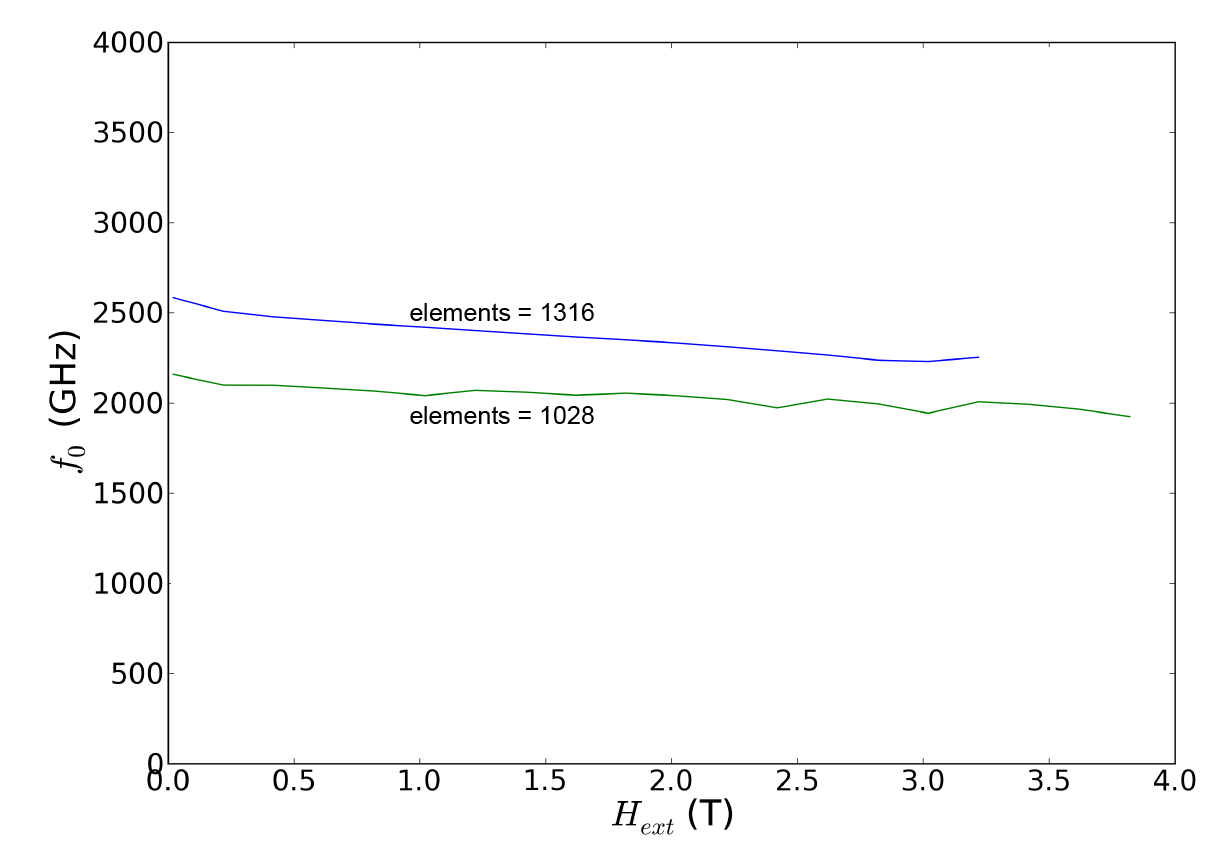}

\caption{\label{fig:singlephase}Attempt frequency of a single phase media
calculated for two different mesh sizes (For the calculation of $\Omega_{0}$
a fitting function from eigenvalue index 30 onwards is used).}

\end{figure}

From an application point of view it is interesting to compare the
attempt frequency of a grain with graded anisotropy to the attempt
frequency of grain for which anisotropy is uniform throughout. Fig.
\ref{fig:singlephase} shows the attempt frequency of a single phase
grain, where all the material parameters are the same as for the graded
media grain of Fig. \ref{fig:graded_fitted} except that the anisotropy
is a constant $K_{1}\sim3.6$MJ/m\textthreesuperior{}. This value
equals the maximum anisotropy constant of the graded media grain of
the previous results.

\section{Results and discussion}

A numerical implementation of the transition state theory was described
with application to two example problems: reversal of a single phase
and graded media magnetic grains. The implementation makes use of
the micromagnetic package FEMME. This method allows for the calculation
of the long term thermal stability of magnetic nanostructures without
any free parameters. The input parameters of the model are the exchange
constant, the anisotropy constant the spontaneous saturation magnetization,
the damping constant and the geometry of the magnet. The advantage
of the presented method over Langevin-dynamic simulation is that the
thermal stability of large magnetic nanostructures can be calculated.
Furthermore Langevin-dynamic simulations are restricted to time scales
of several nano-seconds due to computational effort.

In previous work it was shown that so called exchange spring media
exhibit superior writeability as compared to single phase media, without
lowering the energy barrier \cite{suess_multilayer_2006}. The energy
barrier was assumed to determine the thermal stability. In this paper
we show the exact relation of the energy barrier and the thermal stability
by calculating the attempt frequency. In the present paper we show
that the attempt frequency and energy barrier aare comparable for
a single phase media and a graded media grain with single phase grain
anisotropy is the same as the maximum anisotropy in the graded media
grain. Hence, it was shown that these two media types indeed have
very similar thermal stabilities but the coercive field for the graded
media grain is about a factor of seven smaller than the single phase
grain.

The financial support of the FWF projects P20306-N16 and SFB ViCoM
(F4112-N13) is acknowledged.

\bibliographystyle{plain} \bibliographystyle{plain}
\bibliography{suesslib}

\newpage{}

\pagebreak{}%

\pagebreak{}%

\pagebreak{}%

\pagebreak{}%

\pagebreak{}%

\pagebreak{}%

\end{document}